\documentclass[10pt,aps,longbibliography,prd,superscriptaddress,twocolumn,amsmath,amssymb,floatfix,
nofootinbib %this makes footnotes appear below the page, and not in the bibliography like the revtex class does longbibliography
]{revtex4-1}   
           
\usepackage[dvips]{graphics}
\usepackage{bm}
\usepackage[export]{adjustbox}
\usepackage{epsfig}
\usepackage{enumerate}
\usepackage{subfigure}
\usepackage{color}
\usepackage{graphicx} 
\usepackage[dvipsnames]{xcolor} 
\usepackage{graphicx}
\usepackage[colorlinks,citecolor=blue,linkcolor=blue,urlcolor=blue,plainpages=false,pdfpagelabels]{hyperref}    

\DeclareMathOperator{\sech}{sech}

\usepackage{tcolorbox}
\usepackage{bbold}

\newcommand{\beq}{\begin{equation}}
\newcommand{\eeq}{\end{equation}}
\newcommand{\bea}{\begin{eqnarray}}
\newcommand{\eea}{\end{eqnarray}}

\usepackage{multibib}
\newcites{supp}{Supplementary References}

\begin{document} 
\title{Volkov-Pankratov states in a driven semimetal for a generic interface} 

\author{Aiman Rauf} 
\affiliation{Department of Physics, Jamia Millia Islamia, New Delhi-110025, INDIA}
\author{SK Firoz Islam} 
\affiliation{Department of Physics, Jamia Millia Islamia, New Delhi-110025, INDIA}

\begin{abstract}
Volkov-Pankratov states are nontopological massive bound states which generally arise across the smooth
interface between two adjacent regions of a two-band semimetal, over which a gap parameter changes sign
smoothly. In this work, we show that these modes can be engineered even for a generic smooth interface without
any sign inversion. We consider threefold and twofold topological semimetals in which two adjacent regions are
illuminated by light with different phases. We show that the interface can exhibit an asymmetric Rosen-Morse
potential well for a certain parameter regime even without any sign change of the gap term. Such a quantum well
can host a number of Volkov-Pankratov states. We also note that even in a two-band two-dimensional semimetal
like graphene, the Volkov-Pankratov states can emerge if one induces a momentum shift rather than opening
a gap. Finally, we discuss the transport signatures over those interfacial quantum wells. We note that although
the Ramsauer-Townsend effect appears over the symmetric-type Pöschl-Teller potential well, this effect is absent
over an asymmetric Rosen-Morse potential well. We reveal that a transition from a unit transmission to a unit
reflection can be achieved by just controlling light parameters in a periodically driven graphene. We observe that
the unit reflection phenomenon is direction sensitive; i.e., only incoming electrons from one particular side (left
or right) can be perfectly reflected back without any transmission.
\end{abstract}  
\maketitle
%-------------------------------------------
\section{Introduction}
%-------------------------------------------
The gapless one-dimensional ($1$D) edge modes or two-dimensional ($2$D) surface states with insulating bulk are the hallmarks of $2$D and three-dimensional ($3$D) topological insulators, respectively. Prior to the discovery of such materials, in the early 1980s, Volkov and Pankratov showed that an interface between two regions of a gapless two-band semimetal with mutually inverted mass term gives rise to a number of localized massive bound states \cite{Volkov,pankratov} in addition to the topological states. These massive states are non-topological in nature and dubbed Volkov Pankratov (VP) states \cite{PhysRevB.96.201302}.  Note that the abrupt sign change generally gives rise to a pair of topologically protected chiral modes instead of massive VP states. A number of theoretical works were carried out in recent times that predicted massless $1$D chiral topological modes in $2$D Dirac-type materials with the inverted mass term \cite{PhysRevB.89.085429,PhysRevLett.101.087204,PhysRevB.91.241404}, where the mass term changes sign abruptly. A unidirectional chiral interfacial electromagnetic wave was also predicted in a time reversal symmetry broken Weyl semimetal \cite{PhysRevB.92.115310}. For the smooth interface, the VP states were obtained in photonic graphene with an inverted mass term \cite{PhysRevLett.100.013904}. Recently, a series of theoretical works investigating the VP states in different types of newly emerged electronic systems have been carried out \cite{PhysRevB.96.201302,PhysRevResearch.2.023146,PhysRevB.100.195412,PhysRevB.102.155311,PhysRevResearch.2.023373,PhysRevB.109.235416,PhysRevResearch.6.013193,PhysRevB.95.125306}. Very recently, several experimental observations of the VP states were also reported \cite{vol_exp1,vol_exp2}.  

The last decade has witnessed a sharp increase in the research interest in periodically driven fermionic systems, especially after the discovery of Floquet topological insulator.  The application of light or irradiation to an electronic system can break time reversal symmetry and lead to a non-trivial topological phase-Floquet topological insulator. Following the proposal of photoinduced topological phase transition \cite{PhysRevB.79.081406} and its experimental discovery \cite{lindner2011floquet,peng2016experimental,zhang2014anomalous,wang2013observation}, a series of theoretical investigations were carried out in this direction \cite{PhysRevB.85.125425,PhysRevB.89.235416,PhysRevLett.110.026603, PhysRevB.89.121401, PhysRevB.99.205429, PhysRevB.98.235424, PhysRevB.108.155308,PhysRevB.102.201105}. The Floquet theory \cite{RevModPhys.89.011004} provides a mathematical framework to study a system's evolution over time and the modulated band structure in the presence of a time periodic external perturbation. Periodically driven non-equilibrium electronic systems have diverse application, namely- shining light on the normal region of a Josephson junction can induce a $0-\pi$ phase transition in its transport signatures \cite{PhysRevB.94.165436,PhysRevB.95.201115}, manipulating spin and valley degrees of freedom in different types  of electronic systems \cite{PhysRevB.85.205428,PhysRevB.108.155408,PhysRevLett.116.016802,PhysRevB.84.195408,PhysRevB.98.075422,PhysRevB.90.125438}. A series of theoretical works also revealed the possibility of photo tunability of Weyl nodes \cite{PhysRevLett.117.087402,PhysRevB.94.235137,PhysRevB.97.155152,PhysRevB.94.041409,PhysRevB.96.041205}. 

The Weyl semimetal is a new class of $3$D topological materials that has gained intense research interest in the last decade, because of its unique band structure and transport signatures \cite{Weyl_review1,Weyl_review2}. Its electronic properties are described by the Weyl equation, instead of Schrödinger's equation. In addition to the usual two-band Weyl semimetal, another intriguing class of $3$D topological materials is the recently discovered three-fold topological semimetals \cite{Bradlyn_2016}. In contrast to two-band Weyl semimetals, these  semimetals exhibit a dispersionless flat band in addition to the conical bands. Photo driven band structure modulation and interfacial chiral modes were also studied in recent times in these materials \cite{PhysRevB.100.165302}. In this work, we consider two and three-band topological semimetals and consider an interface between the two adjacent regions  driven by the irradiation with different phases. We note that an interfacial quantum well can emerge for certain parameter regime even without any sign change of the light-induced symmetry breaking parameter. The interface is considered to be smooth which gives rise to the VP states. We obtain exact analytical solutions for the VP states across the smooth boundary in both materials.  We find that a transition from a unit transmission to a unit reflection can occur by suitably controlling the light parameters. The unit reflection over the quantum well is exclusively related to the asymmetric nature of the interfacial potential well which stems from the amplitude imbalance of the photoinduced momentum shifts  in $3$D semimetals or mass term in $2$D semimetals.

The remainder of this paper is organized as follows. In Sec.~\ref{Hamil}, we discuss the low-energy effective Hamiltonian of the threefold semimetal and its energy spectrum. Section~\ref{floquet} discusses the photoinduced band structure modulation of an irradiated $3$D threefold semimetal. The emergence of the VP modes in irradiated threefold $3$D semimetals, two-band semimetals like $3$D Weyl semimetals, and monolayer graphene is presented in Sec.~\ref{vol_pan}. The transport signature, particularly electron transmission over the interfacial quantum well, is given in Sec.~\ref{transport} . A possible experimental setup is described in the Sec.~\ref{expt}. Finally, we summarize in the Sec.~\ref{sum}.

\section{Three-dimensional threefold semimetals}\label{Hamil}
First we consider a gapless threefold $3$D semimetal (TSM) \cite{Bradlyn_2016} that can be regarded as the $3$D counterpart of the $2$D dice lattice \cite{PhysRevLett.112.026402,PhysRevB.84.195422,PhysRevB.96.045418}. One of the notable features of TSMs is the presence of a dispersionless flat band in addition to linearly dispersive conical bands. The low energy effective Hamiltonian for such a material was obtained by using symmetry analysis as \cite{Bradlyn_2016,S_nchez_Mart_nez_2019}
\begin{equation}
H= v\begin{bmatrix}
    0  & e^{i\delta}k_z & e^{-i\delta}k_y \\
    e^{-i\delta}k_z & 0 & e^{i\delta}k_x \\
    e^{i\delta}k_y & e^{-i\delta}k_x & 0 \\
    \end{bmatrix}
\end{equation}
where $v$ is the Fermi velocity, $\delta$ is the real parameter and $c=\hbar= k_B= 1$ is used throughout the calculations. Inside the first Brillouin zone, four such nodal points exist among which we consider only one. The band is non-degenerate at $k\ne0$, unless $\delta=n\pi/3$, where $n$ is an integer. Again for $\pi/3<\delta<2\pi/3$, the Hamiltonian can be adiabatically connected to the one for $\delta=\pi/2$ for which the Hamiltonian can be re-written as $ H=v{\mathbf{k} \cdot S}$. Here, ${{\mathbf{k}}\equiv \{k_x,k_y,k_z\}}$ is the $3$D momentum operator and
${\mathbf S}\equiv\{S_x,S_y,S_z\}$ is the pseudospin-1 matrix given by
\begin{equation}
 S = i\begin{bmatrix}
    0  & e_z & -e_y \\
    -e_z & 0 & e_x \\
    e_y & -e_x & 0 \\
\end{bmatrix}
\end{equation}
where $\{e_x,e_y,e_z\}$ are the unit vectors along the $\{x,y,z\}$ directions. The energy spectrum disperses linearly with a dispersionless flat band, given by $E_{\lambda, \mathbf{k}}=\lambda v |k|, 0$ where $\lambda=\pm$ denote conduction and valence bands and the flat band corresponds to the zero level.

\subsection{Floquet Hamiltonian and energy spectrum of threefold semimetal}\label{floquet}
In this section, we briefly review the band structure modulation of irradiated TSM, based on Floquet theory \cite{RevModPhys.89.011004}, within high-frequency limit. Consider that the TSM is exposed to an external time-dependent perturbation in the form of irradiation or light propagating along the $z$-direction. The irradiation can be described by a vector potential $A(t) = [A_x \sin(\omega t), A_y \sin(\omega t-\phi),0]$, where $\omega$ is the frequency of the irradiation and $\phi$ is the phase. The effect of this potential on the non-perturbed Hamiltonian can be included via canonical momentum as ${\mathbf{k}} \rightarrow {\mathbf{k}} +e{\mathbf{A}}(t)$ where $e$ represents the electron charge. To solve this irradiated Hamiltonian, we employ the Floquet theory which states that the perturbed Hamiltonian exhibits a set of orthonormal solutions $\xi(t) = \chi(t)e^{-i\varepsilon t}$, where $\chi(t) = \chi(t + T)$, $T$ is the field period, and $\varepsilon$ denotes Floquet quasi-energy. The Floquet states $\chi(t)$ can be further expanded as a sum over Fourier components or Floquet side-band indices as $\chi(t) = \Sigma_n \chi_n (t)e^{in\omega t}$. The quasi-energy spectrum of Floquet Hamiltonian can be obtained by diagonalizing it in the basis of Floquet side bands `$n$'. However, an effective Hamiltonian $H_{eff}$ can be obtained by using the Floquet-Magnus expansion within high frequency limit \cite{RevModPhys.89.011004} as $H_{eff} \simeq H + H^{(1)}_F + ...$. Here,  $H^{(1)}_F=[H_{-}, H_{+}]/\omega$ is the first order correction with
\begin{equation}
    H_m= \frac{1}{T} \int_{0}^{T} H(t)e^{-im\omega t} dt
\end{equation}
and $m= \pm1$. Here $H(t)$ is the time-dependent part of the irradiated Hamiltonian. After simplifying, the effective Hamiltonian can be written as $ H_{eff}=H+vS_z\gamma$ where $\gamma= ve^2(2\omega)^{-1} A_xA_y \sin \phi$. The energy spectrum of the effective Floquet Hamiltonian comes out to be $\varepsilon_{\lambda,k}=\lambda v\sqrt{k_x^2 + k_y^2+(k_z + \gamma)^2}$. The external perturbation does not lead to any gap formation, as noted in the case of irradiated graphene \cite{PhysRevB.79.081406}, rather causes a momentum shift along the direction of light propagation. The momentum shift can be controlled by tuning the phase of light, amplitude as well as the frequencies. 
\subsection{Interfacial modes}\label{vol_pan}
\subsubsection{VP modes in $3$D threefold topological semimetals}
\begin{figure}[htbp]
     \includegraphics[width=0.8\linewidth]{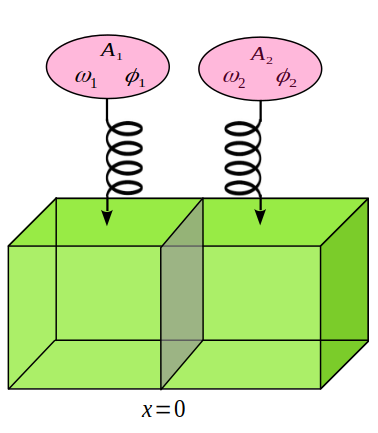}
    \caption{A schematic sketch of the interface in which two adjacent regions are illuminated by two different light sources with different phases ($\phi_1,\phi_2$) is shown.}
    \label{schematic}
\end{figure}

\begin{figure}[t]
 \centering
\begin{minipage}[t]{0.5\textwidth}
\hspace{-.0cm}{\includegraphics[width=.6\textwidth,height=4.5cm]{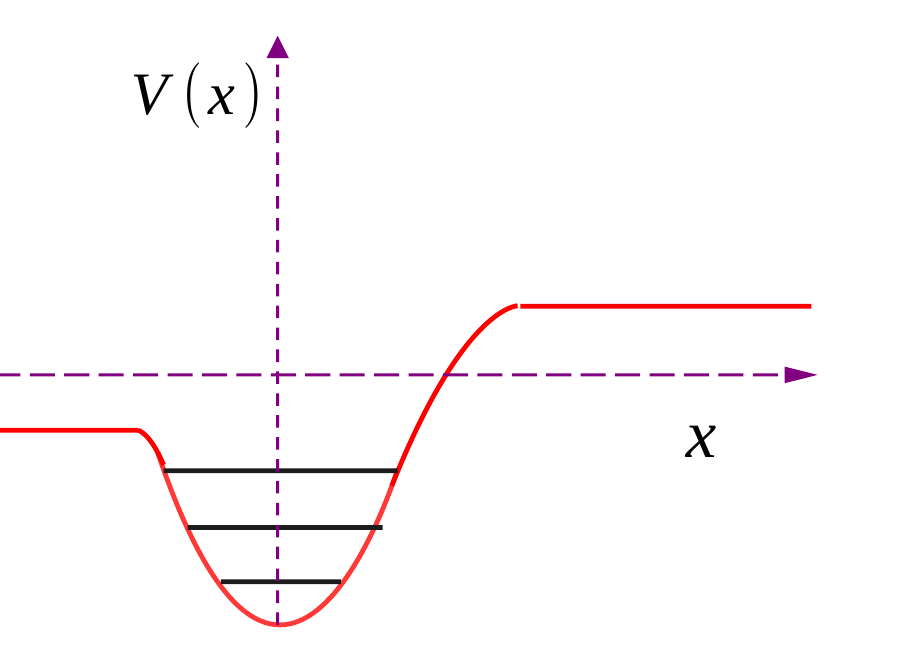}}
\end{minipage}
\caption{A schematic sketch of the Rosen-Morse potential well with few VP bound states are presented.}
 \label{Rosen}
 \end{figure}

To study VP states, we drive the two adjacent regions of TSM by fast oscillating laser fields with different phases as shown in the Fig.~\ref{schematic}, which means that the momentum shifts are not the same in both regions. 
We would like to recall that the interfacial modes for an abrupt interface in $3$D threefold semimetal were studied in Ref.~[\onlinecite{PhysRevB.100.165302}] by one of us, where the photoinduced momentum shift changes sign across the boundary. In contrast, here we consider that the light induced momentum shifts in the two adjacent regions are $\gamma_1$ and $\gamma_2$, respectively i.e., the boundary is very generic and the momentum shifts do not change sign. 
These two regions are separated by a smooth interface of width $L$ over which the momentum shift $\gamma_1$ changes to $\gamma_2$. Let's consider that the interface is in the $y-z$ plane at $x=0$, as schematically shown in Fig.~\ref{schematic}. In such case, the momentum shift can be modelled as $\gamma(x)=\gamma_a+\gamma_r\tanh(x/L)$ where $\gamma_a=(\gamma_1+\gamma_2)/2$ is the average of the momentum shift in the two regions and $\gamma_r=(\gamma_1-\gamma_2)/2$ is the relative difference between them. Hereafter, we set $v=1$. For such smooth interface, after decoupling the eigen value equation $H\Psi(x,y,z)=E\Psi(x,y,z)$ where $\Psi(x,y,z)\sim[\psi_1(x)~\psi_2(x)~\psi_3(x)]^{T}\exp[{i(k_yy+k_zz)}]$, we arrive at
\begin{equation}\label{TF_eigen}
    \left[-\frac{\partial^2 }{\partial x^2} + V_{3f}(x)\right]\psi_2= \tilde{E}^2\psi_2
\end{equation}
where $\tilde{E}=\sqrt{E^2-k_y^2-k_z^2-\gamma_r^{2}}$ and the potential $V_{3f}(x)=-U(E,k_y)\sech^2(x/L)+2\gamma_r(k_z+\gamma_a) \tanh(x/L) $ with $U(E,k_y)= \gamma_r k_y/EL+\gamma_r^2$. This potential is a well-known Rosen-Morse potential \cite{PhysRev.42.210}. There exists a region of momentum satisfying $ U(E,k_y)>0$ and $|\gamma_r(k_z+\gamma_a)|<U(E,k_y)$ for which the potential acts as a well instead of a barrier. The bound states formed inside the well satisfy the following equation
\begin{equation}\label{VP_3fold}
    E^2_n=k_y^2+(k_z+\gamma_a)^2+\gamma_r^2-\frac{\Gamma^2(E,n)}{4L^2} -\frac{4(k_z+\gamma_a)^2L^2\gamma_r^2}{\Gamma^2(E,n)}
\end{equation}
where $\Gamma(E,n)=\sqrt{4U(E,k_y)L^2+1}-(2n+1)$. A sketch of a Rosen-Morse potential well with few bound states inside it is shown in Fig.~\ref{Rosen}.
The above equation is a transcendental equation of energy, which has to be solved numerically to find the bound state solutions. 

 \begin{figure}[t]
 \centering
\begin{minipage}[t]{0.5\textwidth}
\hspace{-.0cm}{ \includegraphics[width=1\textwidth,height=6.5cm]{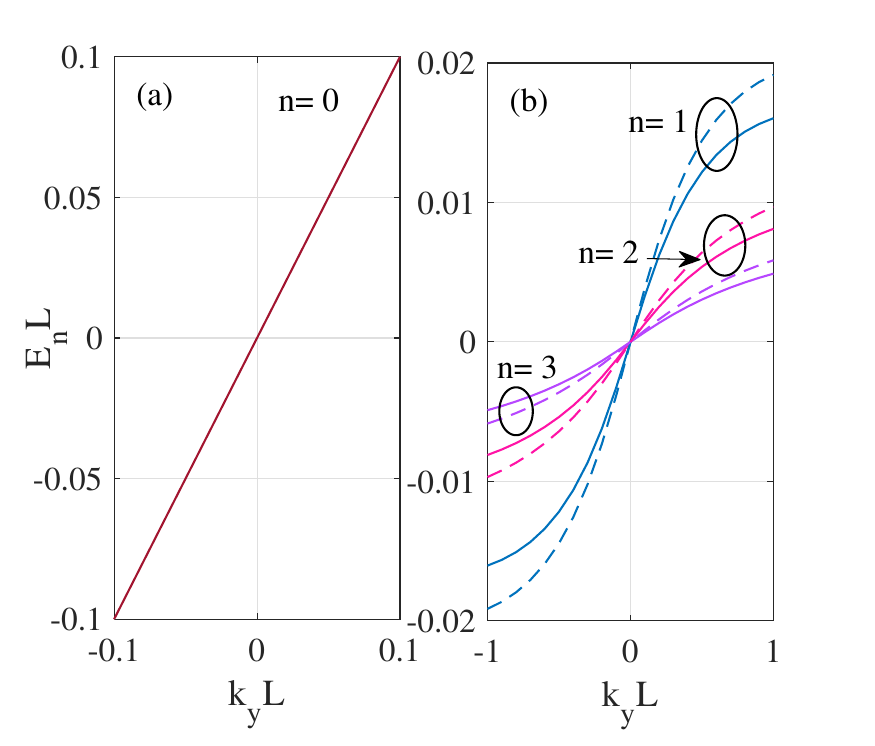}}
\end{minipage}
\caption{(a) The interfacial bound states for $n=0$. This is the gapless unidirectional chiral mode that is topological in nature and does not depend on the nature of boundary details. (b) The dispersion of non-topological VP modes corresponding to $n=1,2,3$. Here, two sets of boundary parameters are used: $(\gamma_aL,\gamma_rL)=(0.3,0.1)$-solid line and $(0.32,0.12)$-dashed-line. These VP modes are significantly affected by slight changes in the boundary details, confirming the non-topological nature. In both cases we set $k_z=0$.}
 \label{VP_states1}
 \end{figure}
The above expression for bound states can be seen to be strongly anisotropic in $k_y-k_z$ space. In order to analyze these modes, we numerically plot first few modes ($n=0,1,2,3$) versus $k_y$ and $k_z$, separately in Fig.~\ref{VP_states1}. First we plot the zeroth ($n=0$) mode using Eq.~(\ref{VP_3fold}) in Fig.~\ref{VP_states1}(a). This is a topological mode which is insensitive to the boundary details i.e., does not depend on the parameters $(\gamma_a,\gamma_r)$. We note that the zeroth mode disperses linearly and is unidirectional in nature.  Note that the topological mode was also obtained for an abrupt interface in Ref.~(\onlinecite{PhysRevB.100.165302}). The VP modes were also discussed there for the special case $k_z=0$. However, in both cases the momentum shift changes sign. On the contrary in the present case the boundary is very generic i.e., momentum shift does not change sign. For example, in Fig.~\ref{VP_states1}, we have taken the boundary parameters as $(\gamma_aL, \gamma_rL)=(0.3,0.1)$ which correspond to $\gamma_1L=0.4$ and $\gamma_2 L=0.2$. Here, the momentum shift $\gamma$ does not change sign rather just smoothly jumps from $\gamma_1L=0.4$ to $\gamma_2L=0.2$. The VP states are also plotted in Fig.~\ref{VP_states1}( b) correspond to $n=1,2,3$. These modes are very sensitive to the boundary parameters, as shown by the solid and dashed lines, which correspond to two sets of boundary parameters: $(\gamma_aL,\gamma_rL)=(0.3,0.1)$ and $(0.32,0.12)$. It is noteworthy to mention that these VP states are also unidirectional and chiral, which is unique contrary to the system like graphene \cite{PhysRevLett.100.013904}.  
 \begin{figure}[t]
 \centering
\begin{minipage}[t]{0.5\textwidth}
\hspace{-.0cm}{ \includegraphics[width=1\textwidth,height=6.5cm]{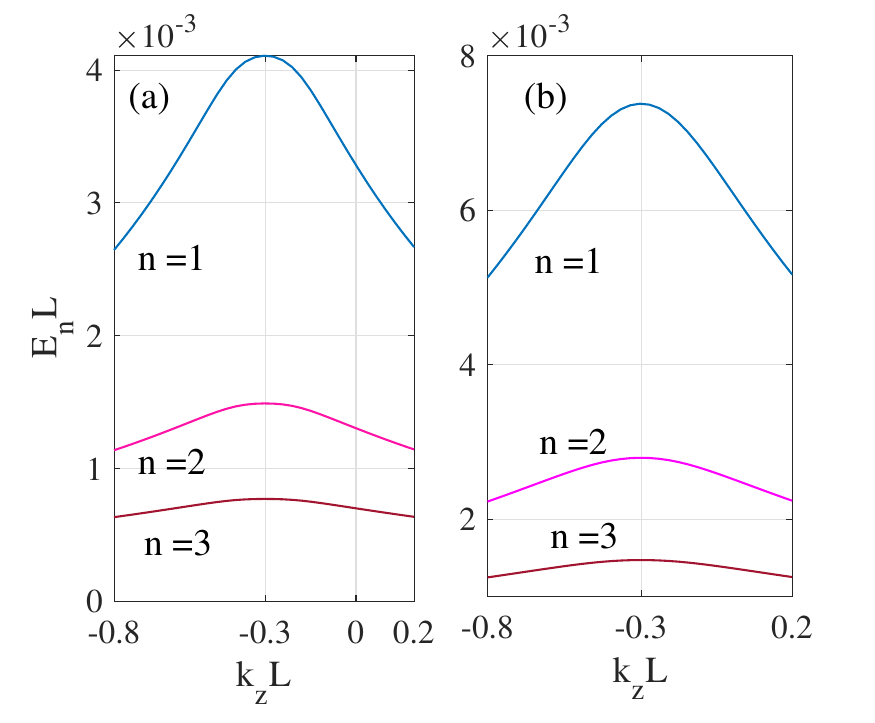}}
\end{minipage}
\caption{The first few VP states are plotted with $k_z$ for (a) $k_yL= 0.1$ and (b) $k_yL= 0.2$, respectively. Here  we keep non-zero $k_y$ in order to widen the range of $k_z$ given by $|k_z|<k_y/E+\gamma_r-\gamma_a$. We also keep the same boundary parameters  as $(\gamma_a L,\gamma_rL)=(0.3,0.1)$. The VP states vary weakly with $k_z$ and exhibit a maximum at $k_z=-\gamma_a$.}
 \label{VP_states2}
 \end{figure}
Note that the number of `n's are limited by the condition $E_n^2>0$ so that the energy is always real. 

The behaviour of the VP states with $k_z$ are also shown in Fig.~\ref{VP_states2} for two sets of $k_y$ as $k_yL=0.1$ and $k_yL= 0.2$. Here $k_y$ is taken to be non-zero just to keep a wide range of $k_z$ corresponding to an interfacial potential well instead of a barrier, as defined by $|k_z|< k_y/EL+\gamma_r-\gamma_a$. We note that VP states are very weakly varying with $k_z$ and the solution is restricted to positive energy only because of the aforementioned condition. We also note that no zero mode solution appears here. It can be understood from the fact that the zero mode is a topological mode and it does not depend on the type of interface. For abrupt interface with inverted momentum shift, an exact solution for zero mode was found in Ref.~(\onlinecite{PhysRevB.100.165302}) as $E_0=sgn(\gamma)k_y$ which has no dispersion with $k_z$ and that is exactly what we obtain for the smooth generic interface in Fig.~\ref{VP_states1}(a). With changing $k_y$, the behaviour remains almost unaffected except a mismatch in the amplitude. However, the lowest VP modes are relatively most sensitive to $k_z$

\subsubsection{VP modes in $3$D twofold topological semimetals}
In this section, we discuss the VP states for a two-band $3$D topological semimetal, namely, the Weyl semimetal. This material has been a focus of the research community recently \cite{Weyl_review1}. The low energy effective Hamiltonian of a twofold $3$D WSM around a particular Weyl node can be taken as $H= {\mathbf \sigma\cdot {k}}$ with linearly dispersive energy eigen values $E_{\lambda,k}= \lambda |k|$, where $\sigma=(\sigma_x,\sigma_y,\sigma_z)$ are Pauli matrices in the orbital space. The effects of a fast oscillatory laser field can be treated by Floquet theorem within high-frequency limit which yields Floquet energy spectrum $\varepsilon_{\lambda, k}=\lambda \sqrt{k_x^2+k_y^2+(k_z+\gamma)^2}$. Note that unlike in the case of threefold semimetal, here the flat band does not exist. Now similar to the threefold semimetal, here also we consider two adjacent regions which are irradiated by laser field with different light parameters for which the light induced momentum shifts are not same. The two regions are separated by a smooth boundary across $x=0$. We again model the change in momentum shift across the boundary in the same fashion as done previously.  After squaring the eigen value equation $H\Psi(x,y,z)=E\Psi(x,y,z)$  with $\Psi(x,y,z)\sim[\psi_1(x)~\psi_2(x)]^{T}\exp[{i(k_yy+k_zz)}]$, we arrive at
\begin{equation}\label{w_eigen}
     \left[-\frac{\partial^2}{\partial x^2} + V_w(x)\right]\psi(x)= \tilde{E}^2\psi(x)
\end{equation}
where $\psi(x)=\psi_1(x)+i\psi_2(x)$ and
\begin{equation}\label{weyl_pot}
  V_w(x)=-\gamma_r\left(\gamma_r+\frac{1}{L}\right)\sech^2\Bigl(\frac{x}{L}\Bigl)+2\gamma_r(k_z+\gamma_a) \tanh\Bigl(\frac{x}{L}\Bigl).
\end{equation}
This potential also mimics the well-known Rosen-Morse potential well for  $|\gamma_r(k_z+\gamma_a)|<\gamma_r(\gamma_r+1/L)$ which for $\gamma_r>0$ places a restriction on $k_z$ as $|k_z|<(1/L+\gamma_r-\gamma_a)$. Note that the region $|k_z|>(1/L+\gamma_r-\gamma_a)$ corresponds to the potential barrier instead of a well. The bound state energy inside this well can be immediately obtained as
%%%%%%%%%%%%%%%%%%%%%% rosen-Morse well for weyl sm %%%%%%
%\begin{figure}[t]
% \centering
%\begin{minipage}[t]{.5\textwidth}
%  \hspace{-.0cm}{ \includegraphics[width=.5\textwidth,height=4.6cm]{weyl_pot1.eps}}
%  \hspace{-0.4cm}{ \includegraphics[width=.5\textwidth,height=4.6cm]{weyl_pot2.eps}}
%\end{minipage}
%  \caption{The Rosen-Morse potential well for (a) $k_z L=0.2$ and (b) $k_z L=0.5$ for $\gamma_a= 0.1$. Any further increase of $\gamma_a$ does not change the figures qualitatively. }
%  \label{RM_well}
% \end{figure}
 %%%%%%%%%%%%%%%%%%%%%%%%%%%%%%%%%%%%%%%%%%%%%
\begin{equation}\label{VP_weyl}
    E^2_n= k_y^2+(k_z+\gamma_a)^2+\gamma_r^2-\left(\gamma_r-\frac{n}{L}\right)^2-\frac{\gamma_r^2(k_z+\gamma_a)^2}{\left(\gamma_r-\frac{n}{L}\right)^2}.
\end{equation}
The zeroth mode ($n=0$) can easily be found to be $E_{0}={\rm sgn}(\gamma_r)k_y$ which is gapless and topological and does not depend on the boundary details. Whereas, the $n=1,2,3..$ modes are sensitive to the boundary width $L$ and non-topological in nature, these modes are purely due to the smooth nature of the boundary-known as the VP modes. Note that only those `$n$'s are allowed for which $E_n^2>0$ i.e., $E_n$ is real.%%%%%%%%%%%%%%%%%%%%%%%%%%%%%%%%%%%%%%%%%%%%%%%%%%%%%%%%%%%%%%%%%%%%%%%%
\begin{figure}[t]
 \centering
\begin{minipage}[t]{.5\textwidth}
  \hspace{-.2cm}{ \includegraphics[width=.5\textwidth,height=5.3cm]{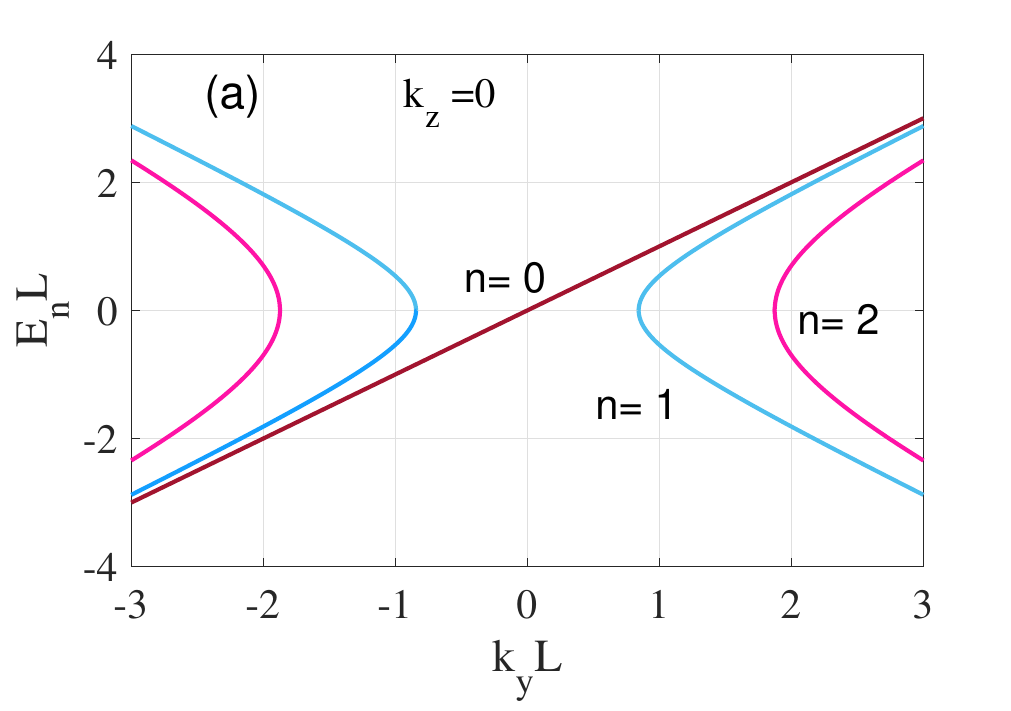}}
  \hspace{-0.5cm}{ \includegraphics[width=.5\textwidth,height=5.3cm]{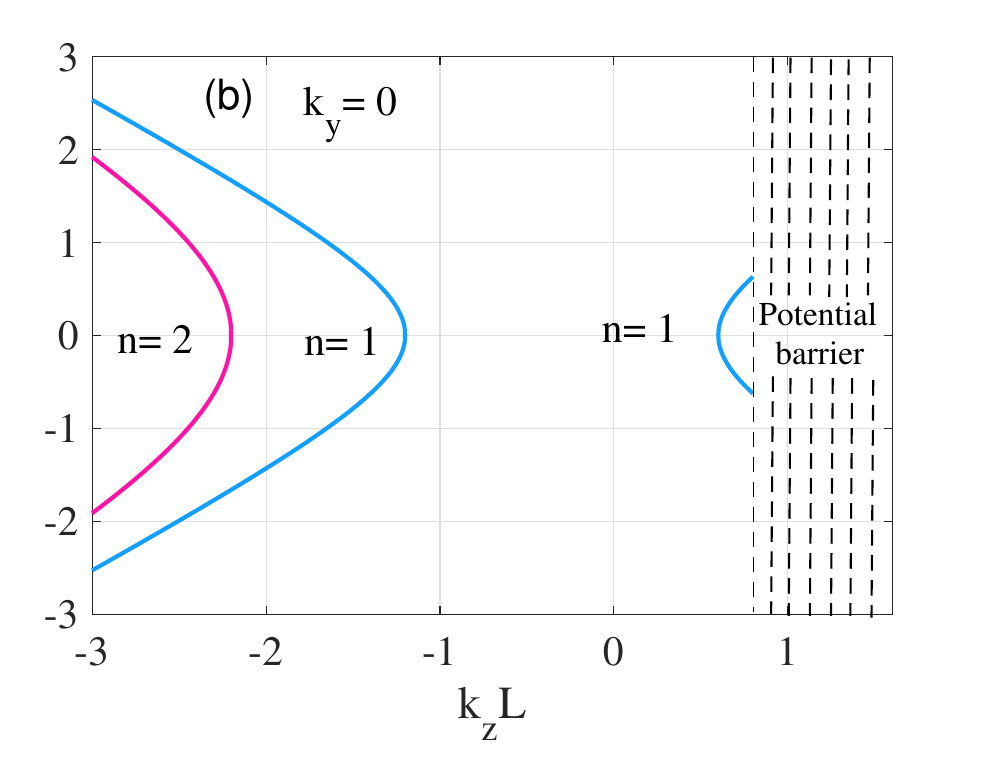}}
\end{minipage}
  \caption{The VP states for (a) $k_z=0$ and (b) $k_y=0$ for $(\gamma_a L, \gamma_r L)= (0.3, 0.1)$. }
  \label{weyl_VP}
 \end{figure}
 %%%%%%%%%%%%%%%%%%%%%%%%%%%%%%%%%%%%%%%%%%%%%%%%%%%%%%%%%%%%%%%%%%%%%%%%
Apart from the zeroth ($n=0$) topological chiral mode, first few VP modes are also shown in the Fig.~\ref{weyl_VP}(a) for $k_z=0$ which are almost hyperbolic in nature and symmetric about $k_y=0$. On the other hand, for $k_y=0$ only $n=1$ mode appears for $k_z>0$ [see Fig.~\ref{weyl_VP}(b)]. However, the  $n>1$ modes do not appear in this region as it does not satisfy the conditions $E_n^2>0$ and $|k_z|<(1/L+\gamma_r-\gamma_a)$ simultaneously, but these higher VP modes are present in the $k_z<0$ region respecting both the aforementioned conditions. Overall, one can see a strong asymmetry in VP modes with respect to $k_z=0$ point.

We can quickly draw a comparison between VP states appearing in the threefold and two-fold Weyl semimetals. Contrary to the case of TSM, here the depth of the interfacial potential well, $\gamma_r(\gamma_r+1/L)$, does not depend on the transverse momentum $k_y$. Additionally, the VP states for Weyl semimetals are not unidirectionally chiral, whereas for TSM they are.

\subsubsection{VP modes in $2$D twofold semimetals: graphene}
\paragraph{Interface between two regions with photoinduced dissimilar mass terms}

Finally, we comment on the generic boundary VP states for the case of $2$D two-band semimetals like graphene. Unlike in the case of $3$D semimetals, application of light propagating along the direction normal to the plane of a $2$D sheet of monolayer graphene opens up a topological gap \cite{PhysRevB.79.081406}. If two neighboring regions of graphene are exposed to the fast oscillating laser field with different phases, two dissimilar mass terms emerge in those regions. Consider that these two mass terms are $\gamma_1$ and $\gamma_2$ and smoothly change from one to another across a boundary of width $L$. We can immediately model the mass term in the same fashion as the momentum shift in the previous cases, as $\gamma(x)=\gamma_a+\gamma_r\tanh(x/L)$. In this case, the interfacial potential well can be obtained as
\begin{equation}\label{mass_term}
 V_g(x)=-\gamma_r\left(\gamma_r+\frac{1}{L}\right)\sech^2\Bigl(\frac{x}{L}\Bigl)+2\gamma_r\gamma_a \tanh\Bigl(\frac{x}{L}\Bigl)
\end{equation}
which is just the $k_z=0$ case for Weyl semimetal as expected, see Eq.~(\ref{weyl_pot}). This potential well exhibits a minimum when $|\gamma_r\gamma_a|<\gamma_r\left(\gamma_r+1/L\right)$ and the bound states inside it can be obtained by just setting $k_z=0$ in Eq.~(\ref{VP_weyl}) as
\begin{equation}\label{VP_gra}
E^2_n= k_y^2+\gamma_a^2+\gamma_r^2-\left(\gamma_r-\frac{n}{L}\right)^2-\frac{\gamma_r^2\gamma_a^2}{\left(\gamma_r-\frac{n}{L}\right)^2}.
\end{equation}
These are the bound state energies which arise inside the interfacial Rosen-Morse potential well for a generic boundary even without any sign change in the mass term. We can quickly check that the zeroth topological mode $E_{0}={\rm sgn}(\gamma_r)k_y$, whereas, $n>1$ modes are non-topological and sensitive to the boundary details-known as the VP modes. If we consider the interface across which the topological mass term changes sign, we can set $\gamma_a=0$ and $\gamma_r=\gamma$ in Eq.~(\ref{VP_gra}) that will return the well-known VP states \cite{PhysRevLett.100.013904}
\begin{equation}
E_n=\pm\sqrt{ k_y^2+2\gamma \frac{|n|}{L}-\frac{n^2}{L^2}}
\end{equation}
provided $(k_yL)^2+2\gamma L|n|>n^2$. Here, the VP states corresponding to different $n$'s exhibit different gaps in the band spectrum.
\paragraph{Interface between two regions with dissimilar momentum shifts}

We have seen that by applying light, a topological gap opens up instead of a momentum shift as in $2$D semimetals. Hence, a question arises, is it possible to engineer VP states in $2$D semimetals without opening any gap? To answer this we consider the application of longitudinal strain instead of light. It was already established that strain can cause a momentum shift which is of course sensitive to the valley index in graphene. The low energy effective Hamiltonian of a strained graphene \cite{PhysRevLett.103.046801} is given by $H=\sigma_xk_x+\sigma_y(k_y-\gamma)$ in one particular valley, where $\gamma$ denotes the strain engineered momentum shift. If two adjacent regions of graphene are strained differently, the momentum shifts will be different in these two regions. The momentum shift can be modelled in the same way as done in the previous sections and the interfacial potential well can be obtained as
\begin{equation}
 V_g(x)=-\gamma_r\Bigl(\gamma_r+\frac{1}{L}\Bigl)\sech^2\Bigl(\frac{x}{L}\Bigl)+2\gamma_r\Bigl(k_y+\gamma_a \Bigl)\tanh\Bigl(\frac{x}{L}\Bigl)
\end{equation} 
which is again a Rosen-Morse potential that acts as a well for $|(k_y+\gamma_a)L|<(\gamma_r L+1)$. The bound state energy inside this well can be obtained as
\begin{equation}
    E^2_n= (k_y+\gamma_a)^2+\gamma_r^2-\left(\gamma_r-\frac{n}{L}\right)^2-\frac{\gamma_r^2(k_y+\gamma_a)^2}{\left(\gamma_r-\frac{n}{L}\right)^2}
\end{equation}
which indicates that the VP states can even arise in $2$D two band semimetal like graphene without the opening of a gap. However, interestingly here we note that the zeroth mode $n=0$ is actually the zero-energy mode $E_0=0$ and does not depend on the width of the boundary. This zero energy mode is localized across $x=0$ and extended along the $y$-direction, for which this mode might be dubbed the zero-line mode. All the other modes  with $n>0$ are sensitive to the width of the boundary and are identified as the VP states. It is noteworthy to mention here that these bound states or the VP modes are robust to weak deformation of the interfacial potential well which can be shown by introducing a deformed hyperbolic function (see Ref.~\cite{PhysRevB.109.235416} for details). 

\section{Transport signature}\label{transport}
 %-------------------------------------------
This section deals with the transmission of electrons through the interface. First, we examine the incoming electron's transmission probability with positive energy over the interfacial potential well corresponding to the inverted mass term, for example, graphene. In this case, the interfacial well becomes Pöschl-Teller potential well instead of Rosen-Morse well. The transmission probability for an incoming electron with momentum $(k_x,k_y)$ with $\tilde{E}^2>0$, over the Pöschl-Teller potential well can be obtained as \cite{landau2013quantum}
 \begin{equation}\label{RT}
 T(k_x,\gamma_r)=\frac{\sinh^2(\pi k_x L)}{\sinh^2(\pi k_x L)+\cos^2[\frac{\pi}{2}(2\gamma_rL+1)]}.
 \end{equation}
 Here we find that an incident electron can fully pass over the well with $T=1$ without undergoing any reflection when $\gamma_r=j$ with $j=1,2,3..$. This is the well-known {\it Ramsauer-Townsend} (RT) effect. As the depth of the well is determined by $\gamma_r(\gamma_r+1)$ which is in fact purely determined by the light parameters, we can conclude that the unit transmission (RT-effect) can be achieved by just simply tuning light parameters externally. The absence of the quantum well is characterised by $j=0$ and hence, it is avoided. The VP index $n$ determines the values of $\gamma_r$ ($\gamma_r=j$) for which the unit transmission takes place. It is noteworthy to mention that similar phenomenon has recently been noted in driven semi-Dirac material \cite{PhysRevB.109.235416}. However, the transmission probability in the semi-Dirac material case depends on the transverse momentum $k_y$ which is the consequence of $k_y$ dependent interfacial quantum well. On the contrary,  the transmission probability $T$ for graphene as seen in Eq.~(\ref{RT}) does not depend on the transverse momentum $k_y$, which enables to fully monitor the  RT effect using external light parameters. 
  
\begin{figure}[htbp]
 \centering
\begin{minipage}[t]{0.5\textwidth}
\hspace{-.0cm}{ \includegraphics[width=.7\textwidth,height=3.5cm]{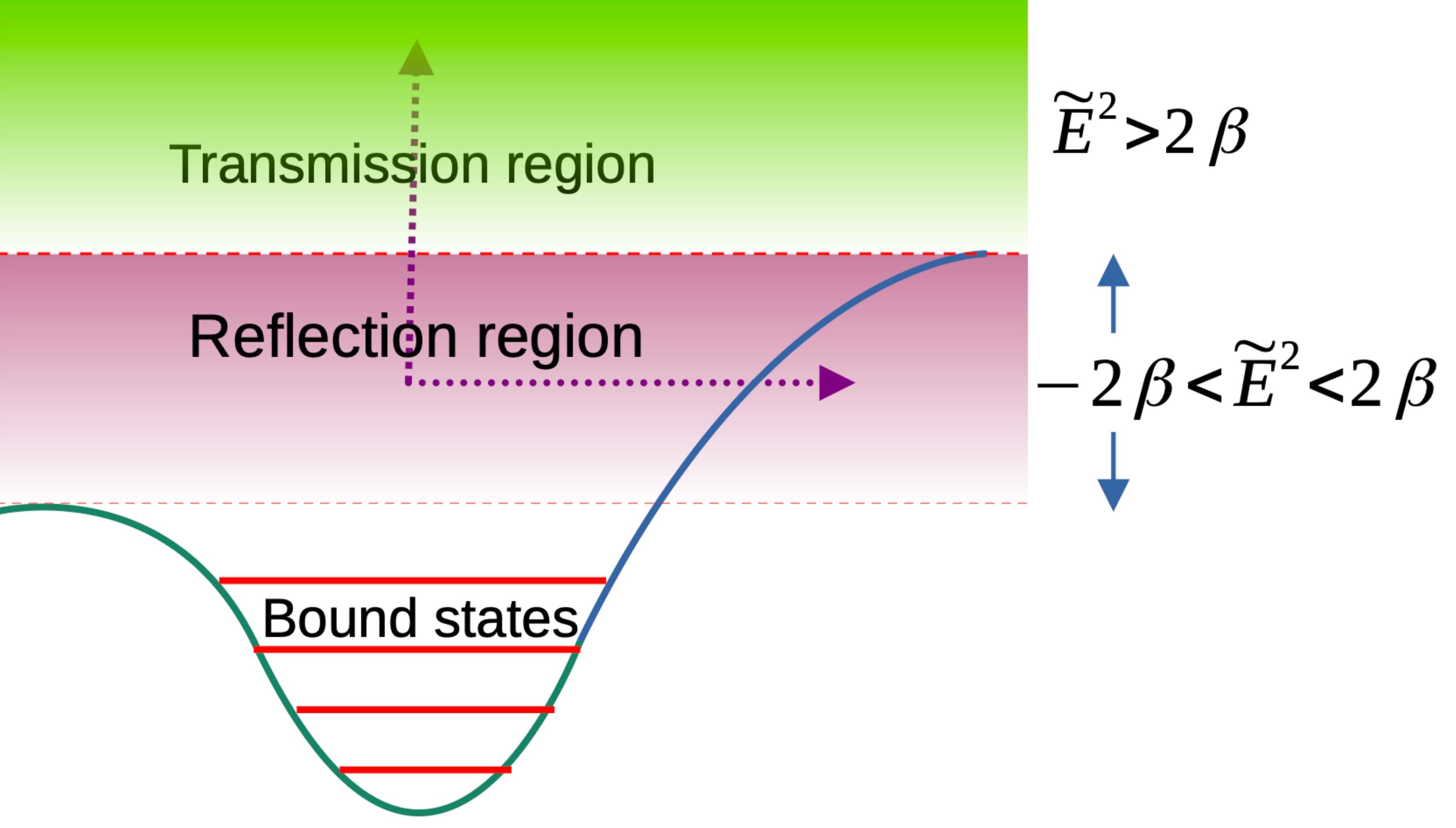}}
\end{minipage}
\caption{Three different regions of scattering are presented schematically for a general Rosen-Morse potential well $V(x)=\alpha(\alpha+1) sech^2(x)-2\beta\tanh(x)$. }
 \label{sketch}
 \end{figure}

 Now we consider the case of a generic interface where the interfacial potential well is not a Pöschl-Teller type but rather a Rosen-Morse quantum well. In such case, three different scattering regions appear as shown in the Fig.~\ref{sketch}, namely the transmission, reflection, and the bound state region \cite{freitas2023generalization}. The transmission region is described by the situation when $\tilde{E}^2>2\beta$ where $2\beta$ is the coefficient of the $\tanh(..)$ term in the Rosen-Morse potential and it measures the degree of asymmetry of the potential well. Note that $\tilde{E}^2=E^2-k_y^2-k_z^2-\gamma_r^2$ is effectively playing the role of energy in the energy eigen-value equations in Eq.~(\ref{TF_eigen}) and Eq.~(\ref{w_eigen}). The transmission region allows electron transmission as well as reflection of incident electron coming from any side, either left or right. The reflection region is described by the regime $-2\beta<\tilde{E}^2<2\beta$, and this region only allows perfect reflection of the incident electron coming from one particular side depending on the sign of $\beta$. The above two regions actually represent a usual $1$D smooth potential step problem with height $2\beta$. The third one is the bound state region when $\tilde{E}^2<-2\beta$, which was already discussed. 
 
 Let us first consider the transmission region for which the
 %%%%%%%%%%%%%%%%%%%%%%%%
 %%%%%%%%%%%%%%%%%%%%%%%%%%
  transmission of an incoming electron over the well can also be obtained as \cite{freitas2023generalization}
  \begin{equation}\label{RM_T}
  T(k_x,\gamma_a,\gamma_r)=\frac{\sinh(\pi k_x L)\sinh(\pi q_x L)}{\sin^2(\pi\gamma_r L)+\sinh^2\left[\frac{\pi L}{2}(k_x+q_x)\right]}
 \end{equation}
 where $q_x=\sqrt{k_x^2-2\beta}$. Note that $\beta$ can take different form for different materials and situation. For example, $\beta=(k_z+\gamma_a)\gamma_r$ when the interface is formed out of unequal momentum shifts, whereas $\beta=\gamma_a\gamma_r$ in graphene when the interface is the result of unequal mass terms. If the mass terms are equal in amplitude but change sign across the boundary then $\beta=0$ as $\gamma_a=0$. In such case, Eq.~(\ref{RM_T}) reduces to Eq.~(\ref{RT}) exhibiting the RT effect at $\gamma_rL=1,2,3..$. Note that in the RT effect, the unit transmission occurs irrespective of the angle of incidence which is in complete contrast to the well-known Klein tunneling phenomenon in graphene \cite{katsnelson2006chiral} where the unit transmission through a rectangular potential barrier occurs at certain angle of incidence.
 
On the other hand for the generic interface when $\gamma_a\ne 0$, the unit transmission is possible only for those incident electrons for which $\beta$ vanishes i.e.,  $k_z=-\gamma_a$ (for graphene $k_y=-\gamma_a$) as can be seen from Eq.~(\ref{RM_T}).  However, it can be seen from the Eq.~(\ref{RM_T}) that for $\gamma_rL=1,2,3..$ and $k_z\ne 0$ the transmission attains a series of maxima instead of unity for the incoming electrons with the $x$-component of the momentum $k_x>\sqrt{2\beta}$. Note that the region $k_x<\sqrt{2\beta}$ corresponds to the reflection region. 
 
 We note that for the case of graphene with non-equivalent mass terms across the interface, the interfacial potential well can be transformed from a symmetric Pöschl-Teller type to an asymmetric Rosen-Morse well, by just switching the light parameters $\gamma_a=0$ to $ \gamma_a\ne 0$ as can be seen from Eq.~(\ref{mass_term}). Such deformation or transition of the interfacial potential well can have interesting signature in the transport phenomenon. The unit transmission occurs at $\gamma_a=0$ and $\gamma_rL=1,2,3..$-the so called RT effect. On the other hand, for the asymmetric case, the transmission can never be unit but a unit reflection can occur for the incoming electrons when energy falls in the reflection region. Hence , we can conclude that a transition can be engineered from a unit transmission to a unit reflection by just controlling the light parameters. This phenomenon resembles the transition from the Klein tunneling to the anti-Klein tunneling of light through an optical analog of the potential step in photonic graphene with respect to the band structure deformation \cite{PhysRevLett.104.063901}. 
  
Note that the unit reflection phenomenon is unidirectional i.e., only the incident electrons coming from one particular side (left or right) can see the potential step and be fully reflected back. However, the directionality of reflection can be reversed by changing the sign of $\beta$ which is in fact possible by controlling the light parameters. In such case, the Rosen-Morse potential well changes to its mirror image about the vertical axis passing through its minima.
 
Now we comment on whether such transition can be achieved  for non-equivalent momentum shifts across the interface.  Let us first consider the case of a two-band Weyl semimetal where the interfacial Potential Well is Rosen-Morse type $V_w(x)$. The transmission over the well can also be described by Eq.~(\ref{RM_T}). The transmission only depends on $k_x$ and not on ($k_y,k_z$). We see that a unit transmission can occur only for those incoming electrons with $k_z=-\gamma_a$, because the interfacial potential well is symmetric for these electrons and no potential step is there to reflect them back. However, for the rest of the incident electrons with $k_z\ne \gamma_a$ the interfacial potential well is the asymmetric Rosen-Morse type and a unit reflection can occur, provided $-2\beta<\tilde{E}^2<2\beta$. Hence, we see that the scenario is not identical to the case of non-equivalent mass terms in graphene. Here a fraction of electrons with $k_z=-\gamma_a$ at $\gamma_rL=,1,2,3$ exhibits unit transmission whereas the unit reflection can occur for the rest of the electrons with $k_z\ne\gamma_a$.

Finally, we comment on the transmission for the threefold semimetal where the interfacial potential well is also the asymmetric Rosen-Morse type as
$V_{3f}(x)$. Here, the transmission over the well can be described by Eq.~(\ref{RM_T}) except that the argument of sine square in the denominator will be modified as it depends on the coefficient of $\sech^2(x/L)$ term in the Rosen-Morse potential well i.e, on  $U(E,k_y)=\gamma_r (k_y/EL+\gamma_r)$ that is energy and $k_y$ sensitive. Hence, we can conclude that unlike in the case of two-band Weyl semimetal, in threefold semimetal the transmission over the interfacial potential well depends on ($k_x$,$k_y$) both. In three-fold semimetals, we also note that for an incoming electron with $k_y=-EL\gamma_r$ the interfacial potential well can transform itself into a smooth potential step $V_{3f}(x)=2\gamma_r(k_z+\gamma_a) \tanh(x/L)$ for which those incoming electrons will be fully reflected back. Interestingly for incoming electrons with $(k_y,k_z)=(-EL\gamma_r,-\gamma_a)$, the interface vanishes.
  
\section{Experimental feasibility}\label{expt}
\begin{figure}[htbp]
     \includegraphics[width=0.7\linewidth, height=4cm]{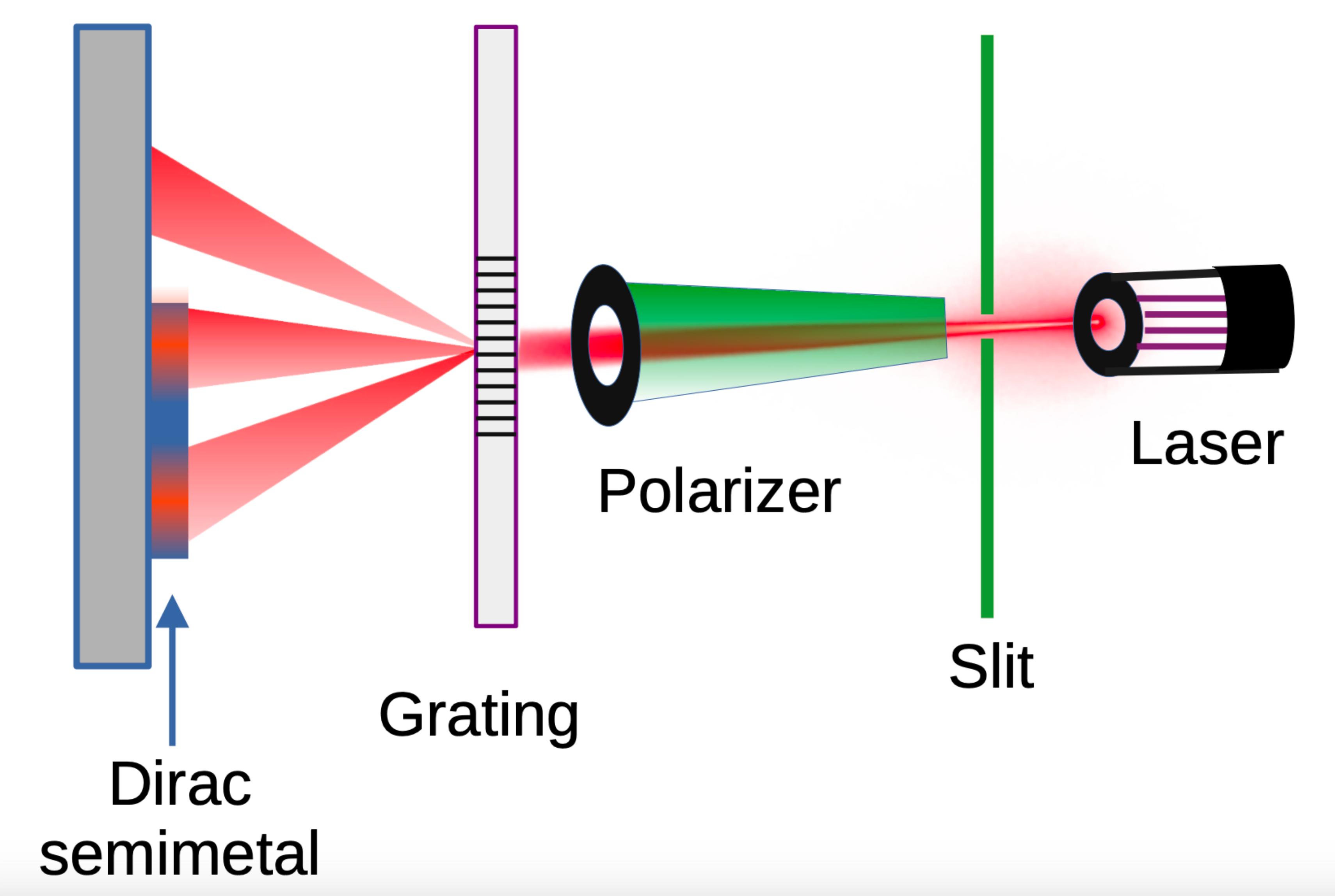}
    \caption{A schematic sketch of the proposed setup illustrating the formation of an interface between two adjacent regions of an electronic system irradiated by two laser beams with different phases.}
    \label{grating}
\end{figure}

Finally, we discuss the experimental feasibility of obtaining VP states.  We need two adjacent regions to be irradiated by light with the same frequency but with different phases. In order to irradiate the two adjacent regions of a semimetal, a single laser beam can be split into two beams with different phases by using a grating and can subsequently  be allowed to be incident on the two adjacent regions of the semimetal. A polarizer can be also placed between the grating and the laser source. The polarized laser beam after passing through the grating forms a diffraction pattern with a number of maxima and minima on the background screen. A graphene sheet or a thick slab of a $3$D semimetal can be placed on the screen in such a manner that two adjacent maxima can cover the entire sample size. The minimum region between the two adjacent maxima can act as a boundary over which the light phase changes smoothly. Because of the path difference between the two adjacent beams, there will always be a phase difference which can be further increased on demand by placing a thin film on the path of one of the two diffracted beams or by controlling the distance between the grating and the surface of semimetals. A schematic sketch of the proposed set-up is presented in Fig.~\ref{grating}. 
Let us quickly comment on two recent experiments \cite{vol_exp1, vol_exp2} on the observation of VP modes. In these experiments, the interface was created out of inverted mass terms, and  was between a trivial insulator and a semimetal, respectively. The signatures of the VP modes were detected in Ref.~(\onlinecite{vol_exp1}) by high frequency compressibility measurements, whereas optical conductivity measurement was performed in the Ref.~(\onlinecite{vol_exp2}). 

Contrary to that, in our case we need photoinduced momentum shifts in $3$D semimetals without any gap opening or gap opening in $2$D semimetal (graphene), without any sign inversion across the interface. The threefold semimetals or triple fermions have been already observed in CoSi \cite{PhysRevLett.122.076402, Nature_Lett567}. On the other hand, a number of materials have been discovered hosting two-band Weyl nodes, see the Ref.~[\onlinecite{Weyl_review2}]. To perturb a quantum system by light, a laser source with the photon energy of about $0.25$ eV with $eA_0=0.01-0.2\AA^{-1}$ can be used. Note that we assume that the wavelength of the incident light is smaller than the interatomic distances in the $3$D Dirac semimetals, for which we can safely ignore the possibility of weak decay of the light amplitude while penetrating inside the bulk of $3$D semimetals. The detection of the VP states can be done by performing an optical transition measurement among different VP states, where an AC electric bias can be used to induce electron excitation among different VP bound states. Here, the chemical potential has to be adjusted by doping accordingly, as the excitation would take place from below to above the chemical potential. A detailed theoretical investigation of the optical signatures of the VP states resulting from a generic interface may be considered later.

\section{Conclusion}\label{sum}
We studied the VP states in two- and threefold topological semimetals for a very generic interface instead of changing the sign of the mass term or momentum shift. In order to induce the momentum shift in $3$D semimetals or the mass term in $2$D semimetals (graphene), we drove the two neighboring regions of the system by a periodically time-dependent perturbation in the form of a fast oscillating laser field with different phases. We noted that a certain momentum or parameter regime exists for which the interface can act as a quantum well and host a number of massive bound states known as VP states. Finally, we also discussed that the interfacial well for the inverted mass term can give a unit transmission for an incoming electron with energy above the potential well: the RT effect. However, for a generic interface, the RT effect does not occur; instead, it exhibits a series of maxima. We also showed that a transition from unit transmission to unit reflection across the interface in graphene can occur by suitably controlling the light parameters. We showed that unit reflection can occur in graphene only if the mass terms on both sides of the interface are unequal. In $3$D semimetals, we note that a fraction of incident electrons can exhibit the RT effect, whereas the rest of the electrons are perfectly reflected back.
\section{Acknowledgments}
The authors acknowledge T. K. Ghosh for useful discussions and valuable comments. 
\bibliography{volkov}
\end{document}